\documentclass[%
showkeys,
 amsmath,amssymb,
 aps,
Prb,
]{revtex4-2}

\usepackage{graphicx}
\usepackage{dcolumn}
\usepackage{bm}
\usepackage{color}
\usepackage{subfigure}

\usepackage{soul}

\begin{document}

\preprint{APS/123-QED}

\title{Elastic Temporal Waveguiding}

\author{Jonatha Santini}
\author{Emanuele Riva}
\email[]{emanuele.riva@polimi.it}
 \affiliation{Department of Mechanical Engineering, Politecnico di Milano, Milano 20156, Italy}

\date{\today}

\begin{abstract}
We provide a theoretical framework to mold time-modulated lattices with frequency conversion and wave-steering capabilities. We initially focus on 1D lattices, whereby a sufficiently slow time-modulation of the stiffness is employed to convert the frequency content of impinging waves. Based on the adiabatic theorem, we demonstrate that undesired reflections, which emerge in time-discontinuous materials, can be eliminated by a careful choice of the modulation velocity. The concept is later explored in the context of 2D lattices, whereby a slow time modulation of the stiffness not only induces frequency conversion without back-scattering, but also serves as a mechanism to steer waves. Our paper explores a new and exciting way to control wave propagation in elastodynamics with scattering-free guiding capabilities, and may open new avenues for the manipulation and transport of information through elastic waves.
\end{abstract}




\maketitle
\section{Introduction}
The dynamics of space, time, and space-time tessellated materials is nowadays explored in several realms of physics to promote rich and unprecendented phenomena not naturally observable. In the context of elastodynamics, materials dressed with space, time, and space-time modulations are conveniently molded to manipulate wave motion at will. Examples have shown different forms of waveguiding that rely on defect mechanisms \cite{casadei2012piezoelectric}, topological insulators \cite{wang2015topological,huber2016topological,miniaci2018experimental,pal2017edge,riva2018tunable,qian2018topology,jin2018robustness,liu2018tunable,riva2020edge}, metagradings \cite{de2021selective,alshaqaq2022programmable,climente2010sound,tol2017phononic,allam20203d,zigoneanu2011design,kim2021poroelastic}, transformation optics and acoustics \cite{pendry2006controlling,leonhardt2006optical,fleury2015invisibility,norris2008acoustic,torrent2008acoustic,xu2020physical,chatzopoulos2022cloaking,quadrelli2021experimental,becker2021broadband,popa2011experimental,colombi2015directional}, which are passive strategies effectively employed to accomplish relevant technological behaviors, such as communication, focusing, and invisibility.

To enrich the dynamical landscape of today, exotic functionalities can be sought by accessing the temporal degree of freedom, which consists of time-varying materials that usually require feeding additional energy into otherwise passive systems. The activation of time-dependent interactions has recently revealed new physics, including studies on unidirectional waves \cite{nassar2020nonreciprocity,trainiti2016non,riva2019generalized,palermo2020surface,park2021spatiotemporal}, adiabatic pumping \cite{xia2021experimental,grinberg2020robust,riva2020adiabatic,riva2021adiabatic}, frequency conversion \cite{yi2017frequency}, and parametric amplification \cite{trainiti2019time}.


All examples above are induced either by space or time heterogeneity, whereby each discontinuity is capable of generating wave scattering or incremental steering. Spatial modulations obey Snell's law, yielding a frequency-invariant wavenumber transformation across the discontinuities. In contrast, a temporal version of Snell's law requires the transformation to occur at constant wavenumber while varying the frequency across the time discontinuity \cite{pacheco2020antireflection}. The complex interplay between consecutive space, time, and space-time modulations, can generate scattering patterns with attenuation, amplification, nonreciprocity, or waveguiding capabilities observable in metamaterials and phononic systems. 
Even though the implementation of space-time and time-discontinuous materials can be challenging, the efforts done in this direction have led to experimental papers which have propelled the research done in the field of active phononics \cite{marconi2020experimental,attarzadeh2020experimental}.

In this paper, we focus our attention on time discontinuous 1D and 2D lattices, providing a link between the time-modulation parameters and the frequency conversion/wave steering capabilities. Motivated by prior works on materials with time-varying permittivity \cite{pacheco2020antireflection,pacheco2020temporal}, we initially pursue frequency conversion in 1D elastic lattices via adiabatic stiffness-modulation. In contrast to Ref. \cite{pacheco2020antireflection}, we herein leverage smooth time-modulations applied to a dispersive elastic medium, whereby the gradual variation of the stiffness is key in avoiding undesired reflections during the frequency transformation of propagating wave-modes (i.e. the states that populate the lattice). The adiabatic theorem is employed to grasp the implied scattering phenomena, and allows us to provide a connection between the modulation velocity and the energy leak between counter-propagating states. In analogy to Ref. \cite{pacheco2020temporal}, this concept is also explored in 2D time-modulated lattices, where the frequency conversion is accompanied by a temporal curvature of wave motion.
We show how this concept can be engineered to produce scattering-free waveguides capable of sending an elastic wave packet between an emitter and a receiver in a controllable manner. 

Our paper explores a new way to manipulate wave motion which may promote additional efforts in the context of time-modulated materials. The procedure employed herein is general and applicable to more complex scenario, such as elastic waves in the continuum, in multi-physics materials, and in more complex metamaterial systems characterized by dispersion relations with remarkable degree of controllability.

\section{Frequency conversion in stiffness-modulated 1D lattices}
\begin{figure}[t]
    \centering
    \subfigure[\label{1Dscheme}]{\includegraphics[width=0.4\textwidth]{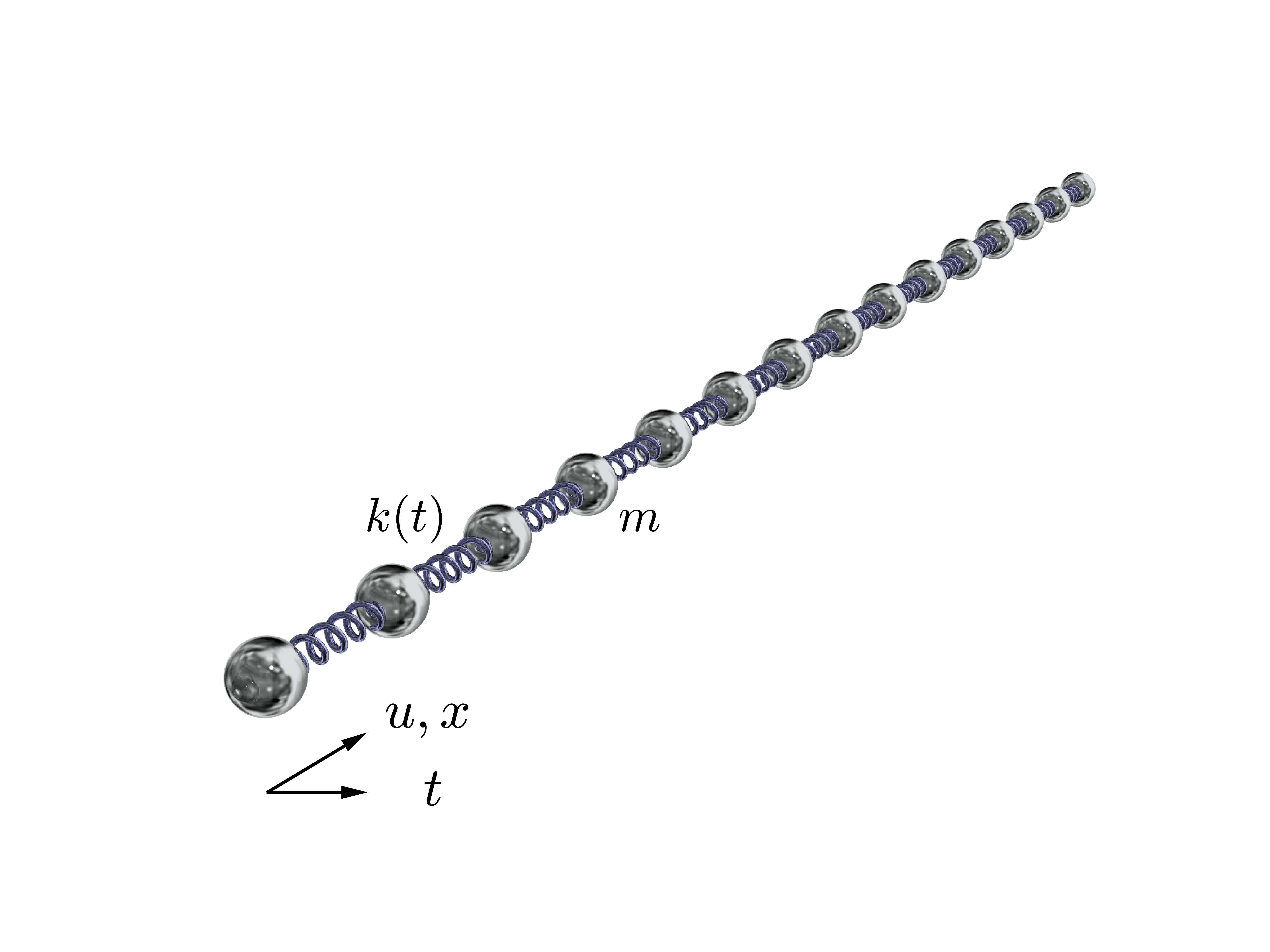}}
    \subfigure[\label{1Dmod}]{\includegraphics[width=0.4\textwidth]{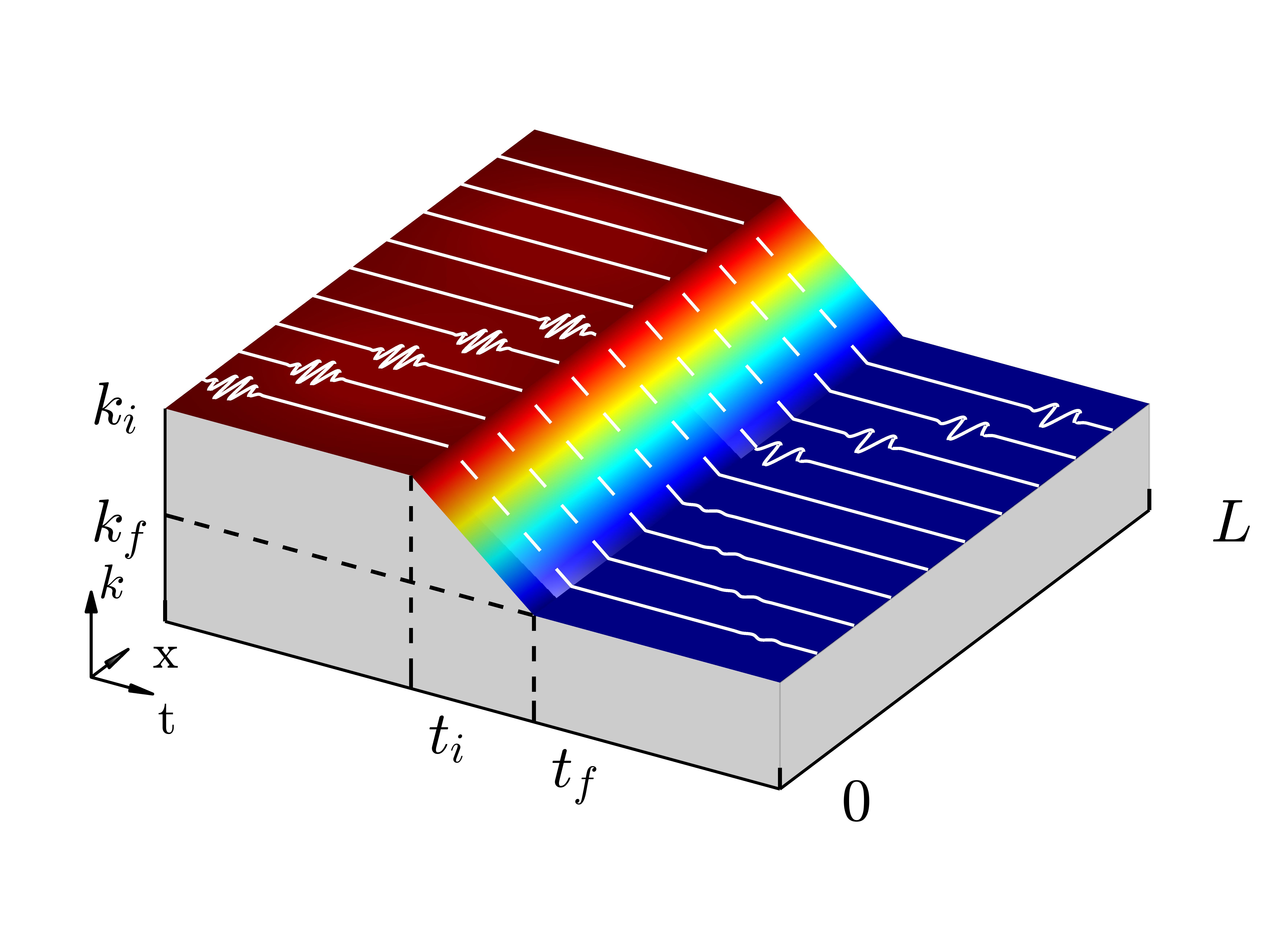}}
    \caption{(a) Schematic of the time-modulated spring-mass system and (b) graphic representation of the time-modulation function, where the scattering and frequency conversion processes are also illustrated with white curves.}
    \label{fig:1D}
\end{figure}
\begin{figure}
    \centering
    \hspace{-.5cm}
    \subfigure[\label{AdPlot.c}]{\includegraphics[width=0.35\textwidth]{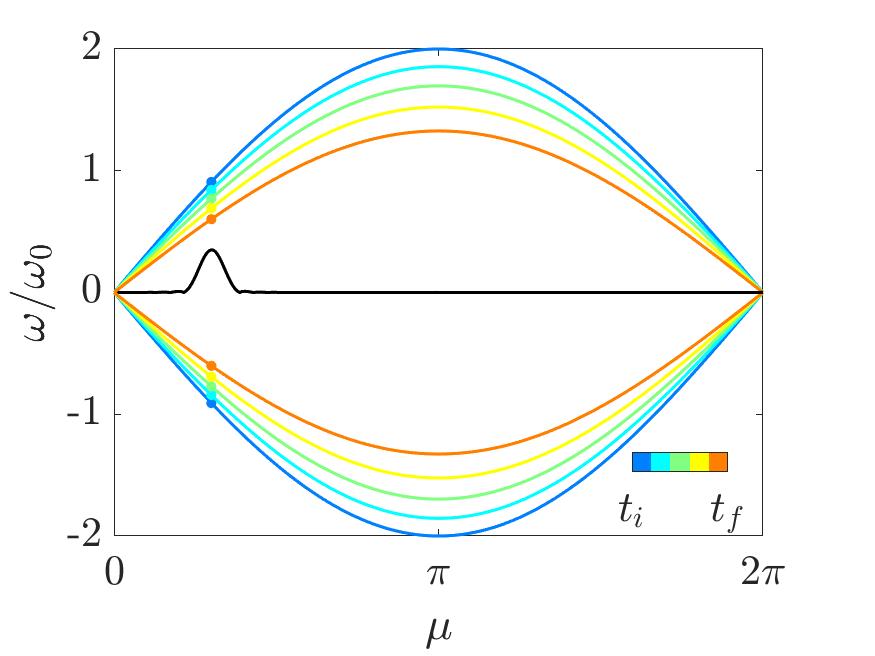}}\hspace{-.5cm}
    \subfigure[\label{AdPlot.a}]{\includegraphics[width=0.35\textwidth]{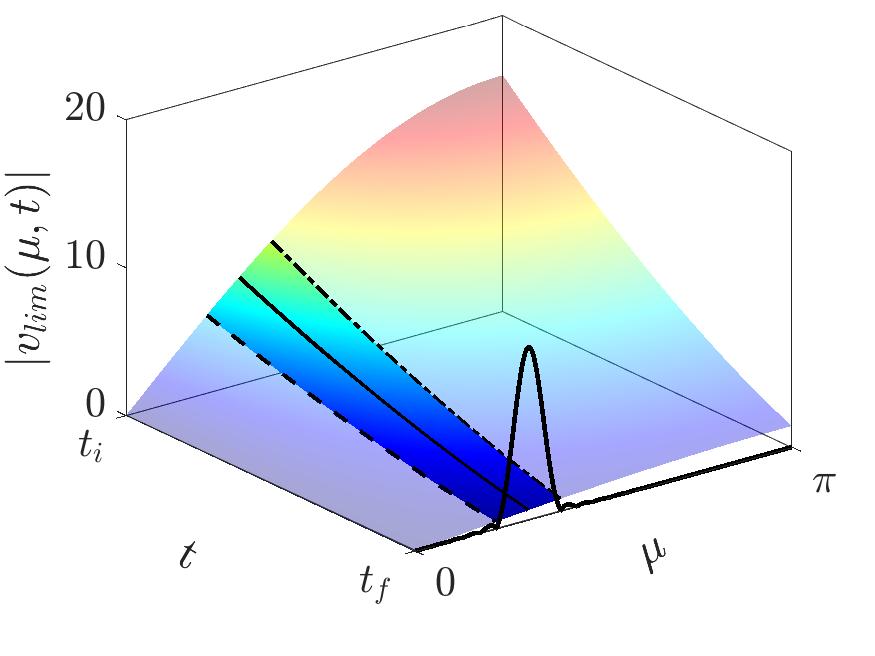}}\hspace{-.25cm}
    \subfigure[\label{AdPlot.b}]{\includegraphics[width=0.35\textwidth]{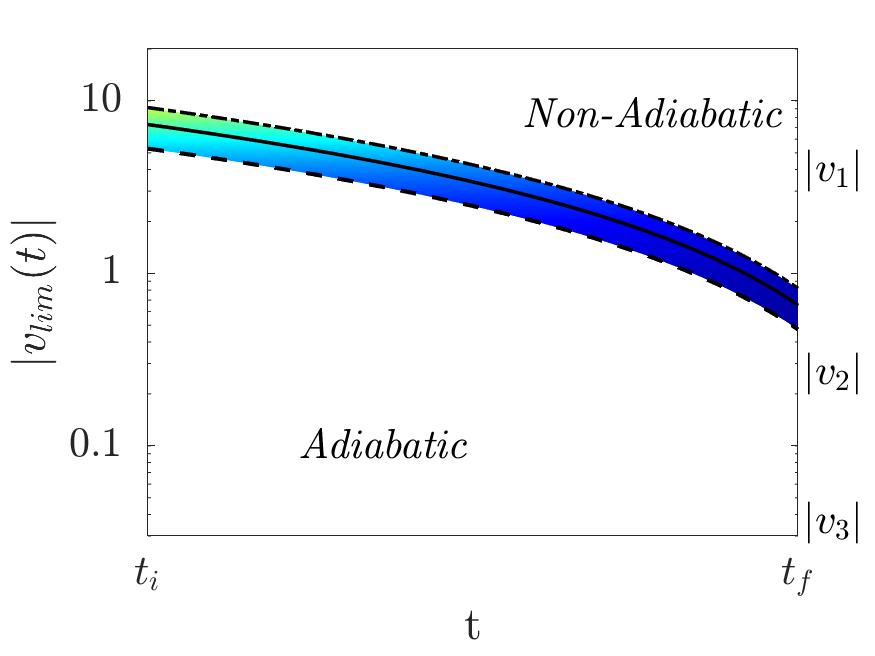}}
    \caption{(a) Dimensionless dispersion relation $\omega\left(\mu\right)/\omega_0$ of the spring mass lattice upon varying time, where $\omega_0=\sqrt{k_i/m}$. Due to stiffness modulation, the dispersion curve fattens. The black curve represents the wavenumber content of the wavepacket employed in the numerical simulations and the colored dots illustrate the frequency conversion that takes place for the central wavenumber of the incident wave packet. (b) Limiting condition for adiabaticity upon varying time $t$ and wavenumber $\mu$. The bright surface highlights the region excited by the incident wavepacket. (c) Side view of the limiting condition for adiabaticity. The three modulation velocities (nonadiabatic $v_1$, intermediate $v_2$ and adiabatic $v_3$) employed in the simulations are indicated alongside the diagram.}
    \label{fig:AdPlots}
\end{figure}

We start the discussion by considering the spring-mass system displayed in Fig. \ref{fig:1D}(a). As mentioned, scope of this section is to perform frequency conversion in such a system without the nucleation of back-scattered waves. We remark that time-modulated materials are known to exhibit frequency conversion and we hereafter present a simple and viable way to pursue this scope.
As such, our lattice is made by a number $N=300$ of unitary masses $m=1$, separated by unitary distance $d=1$, and linked by springs which embody time-varying characteristics $k(t)$. Without any loss of generality, we focus our attention on linear time-modulation laws, but different functions can be employed: 
\begin{equation}
    \begin{split}
        &k(t) = k_i + v(t - t_i)\\
    \end{split}
\label{eq:modLaw}
\end{equation}
here, $v=\left(k_f-k_i\right)/\left(t_f-t_i\right)$ is the modulation speed, which is a fundamental parameter for frequency conversion and allows for controlling the scattering phenomena within the modulated material. $k_i=1$ and $k_f=0.2$ are the initial and final values for the stiffness, $t_i$ is the activation time instant for $k\left(t\right)$ and $t_f$ is the final time of the modulation. 
The key idea is elucidated in Fig. \ref{fig:AdPlots}(b): when a wave impinges across a temporal discontinuity, a pair of counter-propagating waves are generated, whose magnitude depends on the adiabaticity of the temporal transformation and the stiffness-modulation levels. We hereafter provide a qualitative condition that links the reflection and transmission characteristics of the time interface to the velocity of the modulation $v$. 
To that end, we follow and adapt the derivation detailed in our prior works, and we make it applicable to handle wave propagation problems \cite{riva2021adiabatic,xia2021experimental}.

First and foremost, the procedure requires the elastodynamic equation to be expressed by way of a first-order differential form akin to the Schr\"{o}dinger equation. We start the derivation from the second order differential form:
\begin{equation}
    m\ddot{u}_n + 2ku_n - k(u_{n+1}+u_{n-1}) = 0 
    \label{eq:dyn1}
\end{equation}
where $u_n=u_n(t)$ describes the longitudinal motion of the $n^{th}$ mass and, to ease notation, the time dependence of $k$ is implicitly assumed. In the attempt to study the coupling between counter propagating wave modes, we focus on time-dependent dispersion properties of the lattice, which obey the unit cell dynamics with implied Floquet-Bloch conditions. 
Hence, Ansatz $u_n$ are expressed in the complex exponential form $u_n = \hat{u}e^{j\mu n}$, where $\mu = \kappa d$ is the dimensionless wavenumber. After substitution into Eq. \ref{eq:dyn1}, the relevant dynamic equation takes the following convenient form:
\begin{equation}
    \begin{split}
        &|{\hat{\bm\Psi}}\rangle_{,t}=H\left(\mu,t\right)|{\hat{\bm\Psi}}\rangle\hspace{1cm} |\hat{\bm{\Psi}}\rangle=\begin{pmatrix}
        \hat{p}\\
        \hat{u}
        \end{pmatrix}\hspace{1cm}
        H\left(\mu,t\right)=\begin{bmatrix}
        0&-2k(1-\cos\mu)\\[5pt]
        \displaystyle\frac{1}{m}&0
        \end{bmatrix}
    \end{split}
\label{eq:sys}
\end{equation}
where $\hat p=m\hat u_{,t}$ is the linear momentum and $H\left(\mu,t\right)$ is the $2\times2$ time-dependent dynamical matrix with incident wavenumber $\mu$. Note that Eq. \ref{eq:sys} describes the evolution of the amplitude coefficient $|\hat{\bm{\Psi}}\rangle$ for a Bloch wave  $|\hat{\bm{\Psi}}_n\rangle=|\hat{\bm{\Psi}}\rangle{\rm e}^{j\mu}$  propagating through the time-modulated lattice. Due to a temporal version of Snell's law \cite{pacheco2020antireflection}, the impinging wavenumber is invariant during wave propagation and, given a set of initial conditions for $|{\bm\Psi}\rangle$, the solution of Eq. \ref{eq:sys}, effectively describes the evolution of the wave in response to time-modulation, both in terms of amplitude and frequency. This is key for the following discussion.

To shed light on this matter, we consider a waveguide with a quasi-static variation of the properties, i.e. whose modulation velocity is much slower than the time-scale of the underlying dynamics, whereby the solution can be synthesized in a complex exponential form $|\hat{\bm{\Psi}}\rangle=|\bm{\psi}\rangle{\rm e}^{j\omega t}$. As such, for an incident wavenumber $\mu$, the system supports a pair of states $\omega_{1,2}=\pm\omega$ that instantaneously populate the waveguide at time $t$ and are dictated by the right eigenvalue problem $H\left(\mu,t\right)\left|\bm{\psi}^R\right>=j\omega\left|\bm{\psi}^R\right>$. Such states correspond to counter propagating waves whose frequencies are illustrated through the dispersion relations in Fig. \ref{fig:AdPlots}(a). Now, when the temporal modulation is sufficiently slow, a quasi-static evolution takes place such that the impinging wave, either at $\omega_1$ or at $\omega_2$, is frequency-converted consistently to the time-evolution of the dispersion relation marked with colored dots. 
It is now assumed that a right traveling wave packet is present before time modulation, whereby its spectral content is schematically illustrated with the black continuous line in Fig. \ref{fig:AdPlots}(a-b). At this step, we remark that Eq. \ref{eq:sys} describes the time evolution of the amplitudes associated to an arbitrary incident wavenumber $\mu$ and hence, is also suitable to qualify the scattering of non-monochromatic wave-packets. In other words, any incident wave packet can be described through linear combination truncated to $2M+1$ terms in Fourier space $|\bm{\Psi}_{n}\rangle=\sum_{m=-M}^{M}|{\bm{\hat \Psi}}\rangle{\rm e}^{jn\mu_m}$, where $\mu_m=m\frac{2\pi}{Nd}$ are the impinging wavenumbers considered in the expansion and $|{\bm{\hat \Psi}}\rangle$ are the amplitude coefficients which accommodate both momentum and displacement. Note that the amplitude coefficients $|{\hat {\bm{\Psi}}}\rangle$ are effectively the initial conditions at $t=0$ of the problem described in Eq. \ref{eq:sys} and the time evolution of different amplitude coefficients $|{\hat {\bm{\Psi}}}\rangle$ is in general variable with $\mu$. We now analyze the time evolution of an arbitrary incident wavenumber with unitary amplitude and, after that, we generalize to linear combination of them. 

We therefore assume a single value for $\mu_m=\mu$ and we consider a right-traveling plane wave with unitary amplitude coefficient $|{\bm{\hat \Psi}}\rangle=|\bm{\psi}_1^R\rangle{\rm e}^{j\omega_1t}$ as initial condition, where $|\bm{\psi}_1^R\rangle$ is the right eigenvector relative to $\omega_1$.
During time modulation, the solution is synthesized as a linear combination of the two states supported by the waveguide $|{\bm{\hat \Psi}}\rangle=\sum_{n=1}^{2}a_n|\bm{\psi}^R_n\rangle{\rm e}^{j\omega_nt}$, weighted by participation factors $a_n$. In other words, if some energy is present in one state, say $\omega_1=+\omega$, a part of that can leak to $\omega_2=-\omega$, depending on the time-modulation parameters. 

\begin{figure}
    \centering
    \subfigure[\label{fast_a}]{\includegraphics[width=0.3\textwidth]{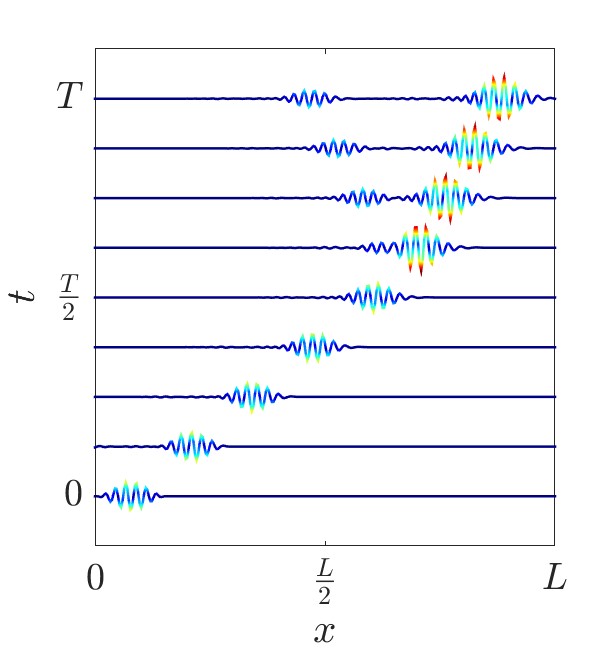}}
    \subfigure[\label{fast_b}]{\includegraphics[width=0.3\textwidth]{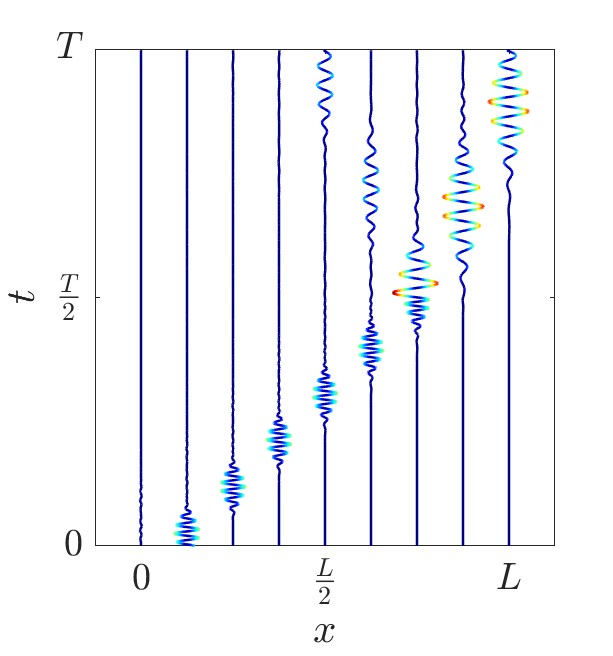}}
    \subfigure[\label{fast_c}]{\includegraphics[width=0.3\textwidth]{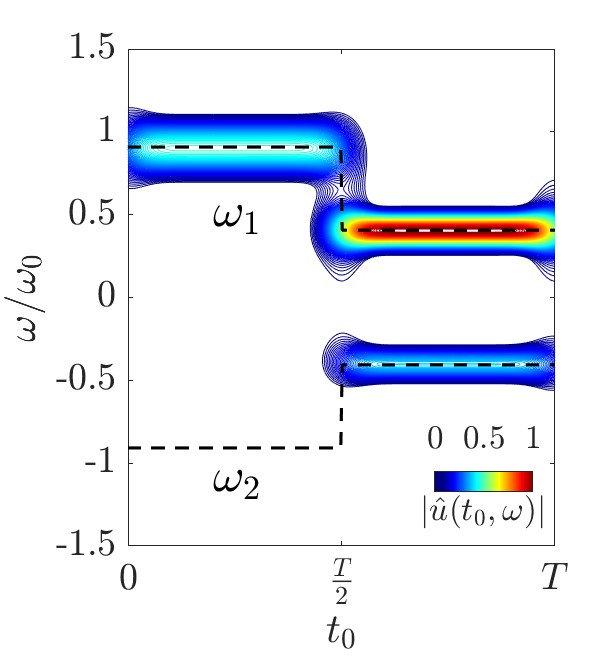}}\\
    \subfigure[\label{medium_a}]{\includegraphics[width=0.3\textwidth]{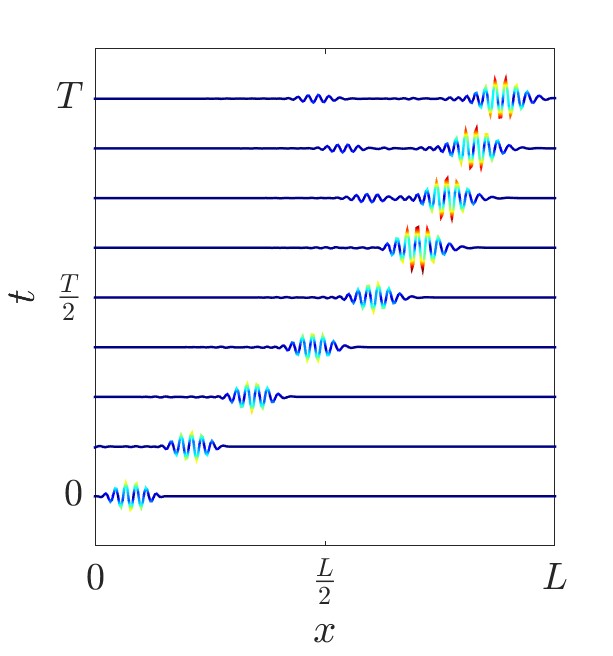}}
    \subfigure[\label{medium_b}]{\includegraphics[width=0.3\textwidth]{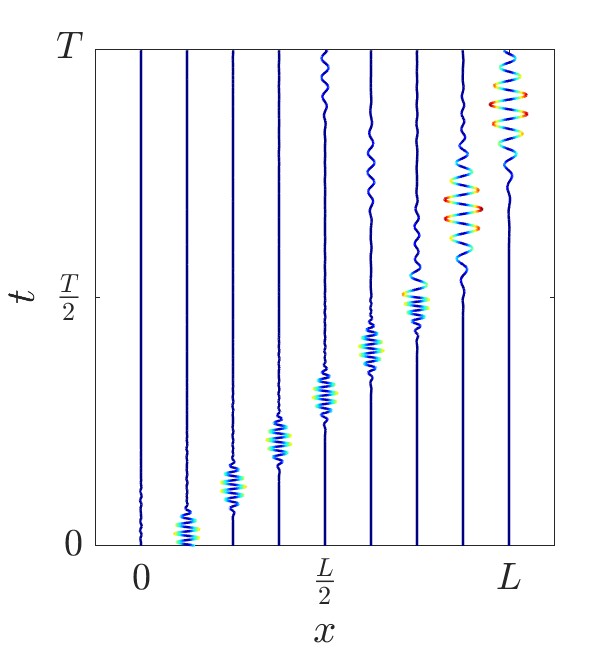}}
    \subfigure[\label{medium_c}]{\includegraphics[width=0.3\textwidth]{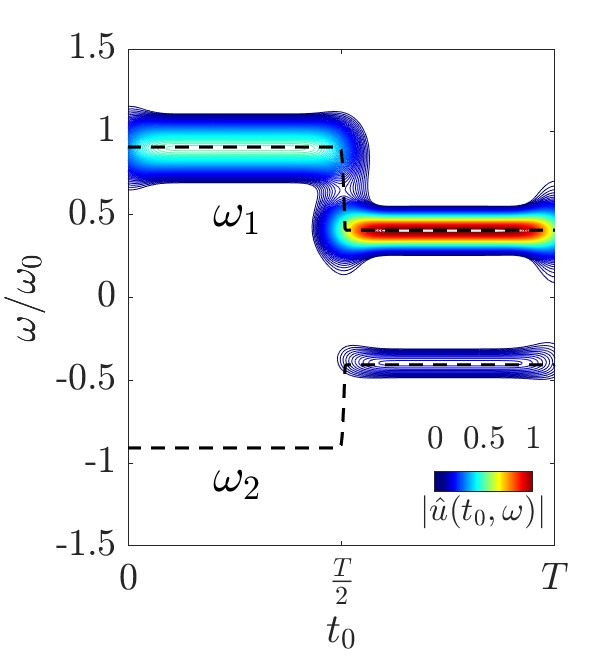}}\\
    \subfigure[\label{slow_a}]{\includegraphics[width=0.3\textwidth]{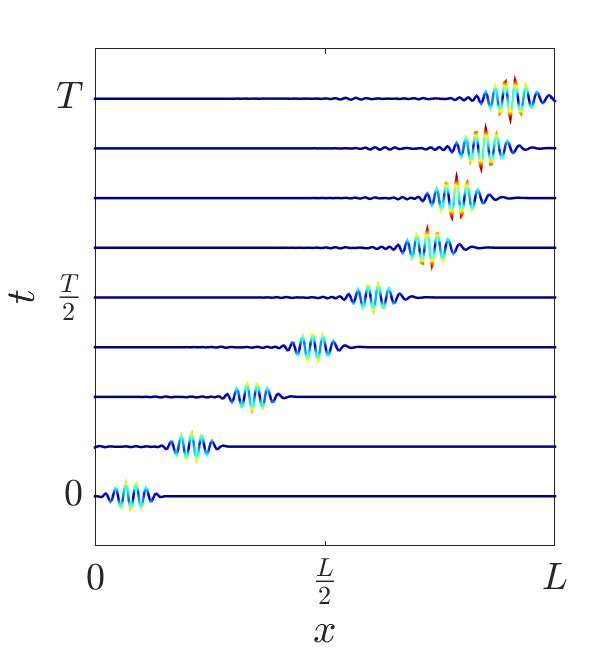}}
    \subfigure[\label{slow_b}]{\includegraphics[width=0.3\textwidth]{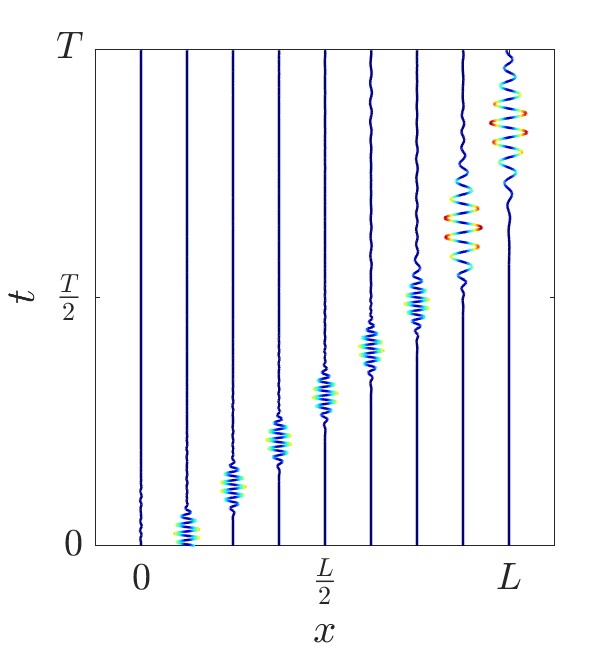}}
    \subfigure[\label{slow_c}]{\includegraphics[width=0.3\textwidth]{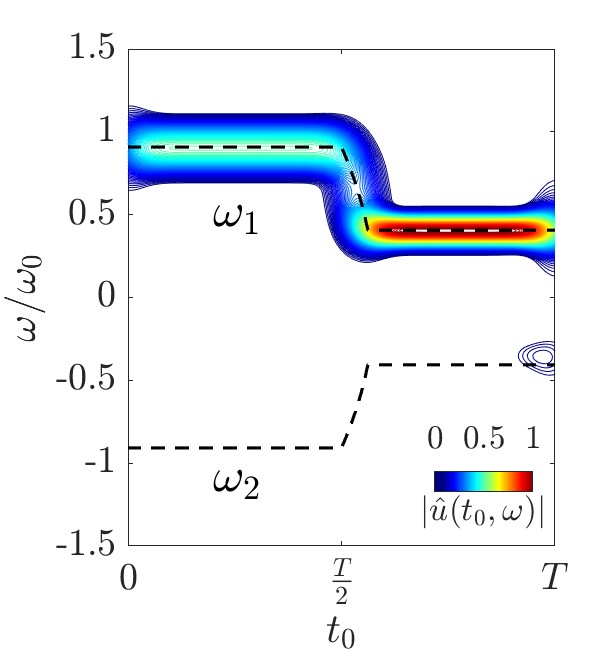}}\\
    \caption{Waterfall plots and corresponding spectrograms for different modulation velocities (a-c) $v_1 = 3.5$, (d-f) $v_2 = 0.23$, (g-i) $v_3 = 0.03$. The plots (a), (d), and (g) represent spatial snapshots of the time history, to highlight the wavenumber-invariant transformation due to time modulation. The plots (b), (e), and (h) are the time histories at fixed positions, to highlight the temporal variation of frequency content. The spectrograms (c), (f), and (i) illustrate the Fourier transformed version  of the displacement field (colored plot) superimposed to the expected temporal evolution of the eigenvalues. }
    \label{fig:dynRes}
\end{figure}

At this step, and following the procedure described in Ref. \cite{riva2021adiabatic,xia2021experimental}, we define a condition for the modulation to provide negligible energy transfer from state $1$ to state $2$, which \textit{de facto} is a condition for adiabaticity:
\begin{equation}
\left|\displaystyle\frac{\left<\bm{\psi}_2^L\left|H(\mu,t)_{,t}\right|\bm{\psi}_1^R\right>}{\left(\omega_2-\omega_1\right)^2}\right|<<1    
\label{eq:limit}
\end{equation}
where, in order for the expression to apply, the left and right eigenvectors $\langle\bm{\psi}_{1,2}^L|$ and $|\bm{\psi}_{1,2}^R\rangle$ are normalized such that $\left<\bm{\psi}_s^L|\bm{\psi}_r^R\right>=\delta_{sr}$. The analytical expressions for the eigenstates are:
\begin{equation}
    \begin{split}
        &\omega_{1,2}=\pm\sqrt{\frac{2k\left(\cos{\mu}-1\right)}{m}}\hspace{1cm}\textit{with:}\\[5pt]
        &|\psi^R_1\rangle = 
        \begin{pmatrix}
            j\sqrt{mk(\cos{\mu}-1)}\\[6pt]
            \displaystyle\frac{\sqrt{2}}{2}
        \end{pmatrix}
        \hspace{1cm}
        \langle\psi^L_1| = 
        \begin{pmatrix}
            \displaystyle\frac{1}{j\sqrt{4mk(\cos{\mu}-1)}}&  \displaystyle\frac{\sqrt{2}}{2}
        \end{pmatrix}
        \\[5pt]
        &|\psi^R_2\rangle = 
        \begin{pmatrix}
            -j\sqrt{mk(\cos{\mu}-1)}\\[6pt]
            \displaystyle\frac{\sqrt{2}}{2}
        \end{pmatrix}
        \hspace{20pt}
        \langle\psi^L_2| = 
        \begin{pmatrix}
            -\displaystyle\frac{1}{j\sqrt{4mk(\cos{\mu}-1)}}&
            \displaystyle\frac{\sqrt{2}}{2}
        \end{pmatrix}
    \end{split}
    \label{eq:modes}
\end{equation}
Merging Eqs. \ref{eq:limit}-\ref{eq:modes}, we get to a limiting condition for the rate of change $v$, in order for the modulation to induce an adiabatic transformation to the incident state:
\begin{equation}
    \left|v_{lim}\right| << 16\left|\sin{\frac{\mu}{2}}\right|\sqrt{\frac{k^3}{m}}
    \label{eq:limSpeed}
\end{equation}
which is function of both incident wavenumber $\mu$ and time $t$. This condition is mapped in Fig. \ref{fig:AdPlots}(b), where the incident wavenumber content, hereafter employed for simulations, is highlighted.

We now perform numerical simulations to qualify the dynamic behavior discussed above, and in the attempt to capture limiting conditions expressed in Eq. \ref{eq:limSpeed}. The values employed in the simulations are $v_{1} = -3.5$, $v_{2} = -0.23$ and $v_{3} = - 0.03$, which are represented in Fig. \ref{fig:AdPlots}(c) and superimposed to the lateral view of Fig. \ref{fig:AdPlots}(b). 
These values correspond to a nonadiabatic, intermediate, and adiabatic modulation speeds for the entire incident wavenumber spectrum from $t_i$ to $t_f$.

Fig. \ref{fig:dynRes} illustrates results for an impinging Gaussian wavepacket with number of periods $n=10$ and central wavenumber $\mu = 0.3\pi$, which is provided as initial condition for the simulations. The time histories are conveniently represented with space-time diagrams in Figs. \ref{fig:dynRes}(a-b), \ref{fig:dynRes}(d-e), and \ref{fig:dynRes}(g-h) from $t=0$ to $t=T$. Specifically, Figs. \ref{fig:dynRes}(a), \ref{fig:dynRes}(d), and \ref{fig:dynRes}(g) illustrate a wavenumber-invariant transformation, whereby the fixed-time displacement of the wave does not change shape during time modulation. 
In contrast, Figs. \ref{fig:dynRes}(b), \ref{fig:dynRes}(e), and \ref{fig:dynRes}(h) suitably represent the frequency change dictated by time-modulation, whereby the temporal period of the wave packet is clearly increased during modulation. Further relevant comments follow: (i) the wave packet decelerates during the transformation, consistently with the dispersion properties displayed in Fig. \ref{fig:AdPlots}(a). (ii) The frequency and velocity changes are accompanied by reflection and transmission of the impinging wave, whose magnitudes are dependent upon the modulation speed. The scattering level is reduced for lower velocity values $v$ and eventually eliminated in case of purely adiabatic modulations. 

This concept is better described by way of Figs. \ref{fig:dynRes}(c), \ref{fig:dynRes}(f), and \ref{fig:dynRes}(i), where the energy content within the waveguide (colored contours) is superimposed to the expected evolution of the states (dashed lines). Such spectrograms are herein evaluated by windowing the time history $u\left(x_n,t\right)$ through a moving Gaussian function $G={\rm e}^{-\left(t-t_0\right)^2/2c^2}$, where the parameter $c=T/20$ controls the width of the window and $t_0$ is smoothly varied between $t_0\in\left[0,T\right]$, to capture the variation of the spectral content in time. The Fourier transformed version of the windowed displacement field $\hat u\left(t_0,\omega,\mu\right)$ is further post-processed by taking the $L^{2}$ norm in wavenumber space limited to $\mu>0$, in order to eliminate one dimension and get to the convenient form $\hat u\left(t_0,\omega\right)$. 
In the figures, the energy is initially stored in the state $\omega_1=+\omega$. After the modulation takes place, a certain amount of energy is scattered toward backward propagating states, i.e. in correspondence of $\omega_2=-\omega$. This energy leak is observable in Fig. \ref{fig:dynRes}(c) and in Fig. \ref{fig:dynRes}(f) with minor intensity. In contrast, the configuration displayed in Fig. \ref{fig:dynRes}(i) exhibits a clear frequency conversion without back-scattering, whereby all the energy remains confined in correspondence of $\omega_1=\omega$, and the transformation can be considered adiabatic. 

\section{Temporal steering in stiffness-modulated 2D lattices}
\begin{figure}
    \hspace{-0.5cm}
    \subfigure[]{\includegraphics[width=0.34\textwidth]{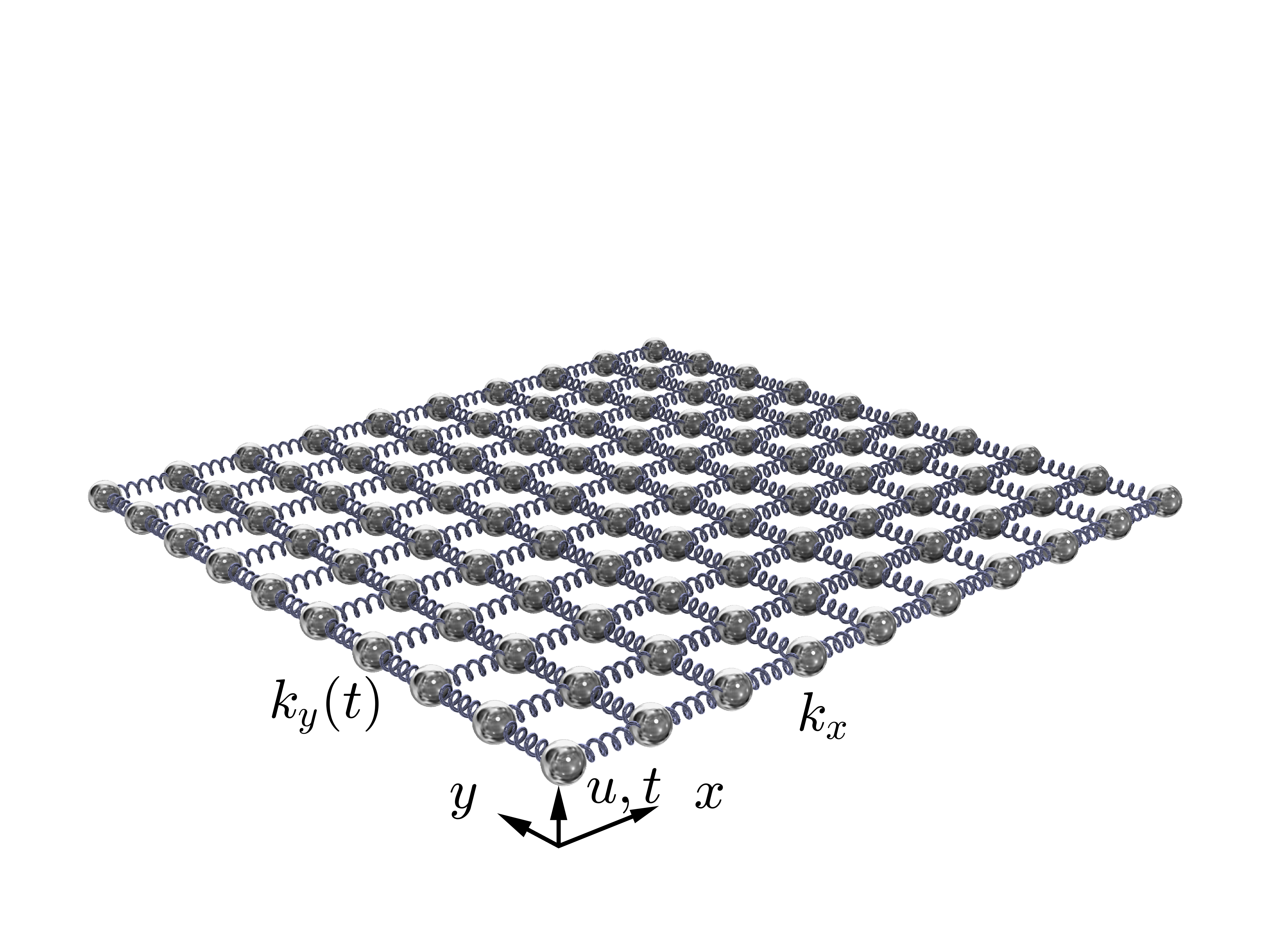}}\hspace{-0.5cm}
    \subfigure[]{\includegraphics[width=0.34\textwidth]{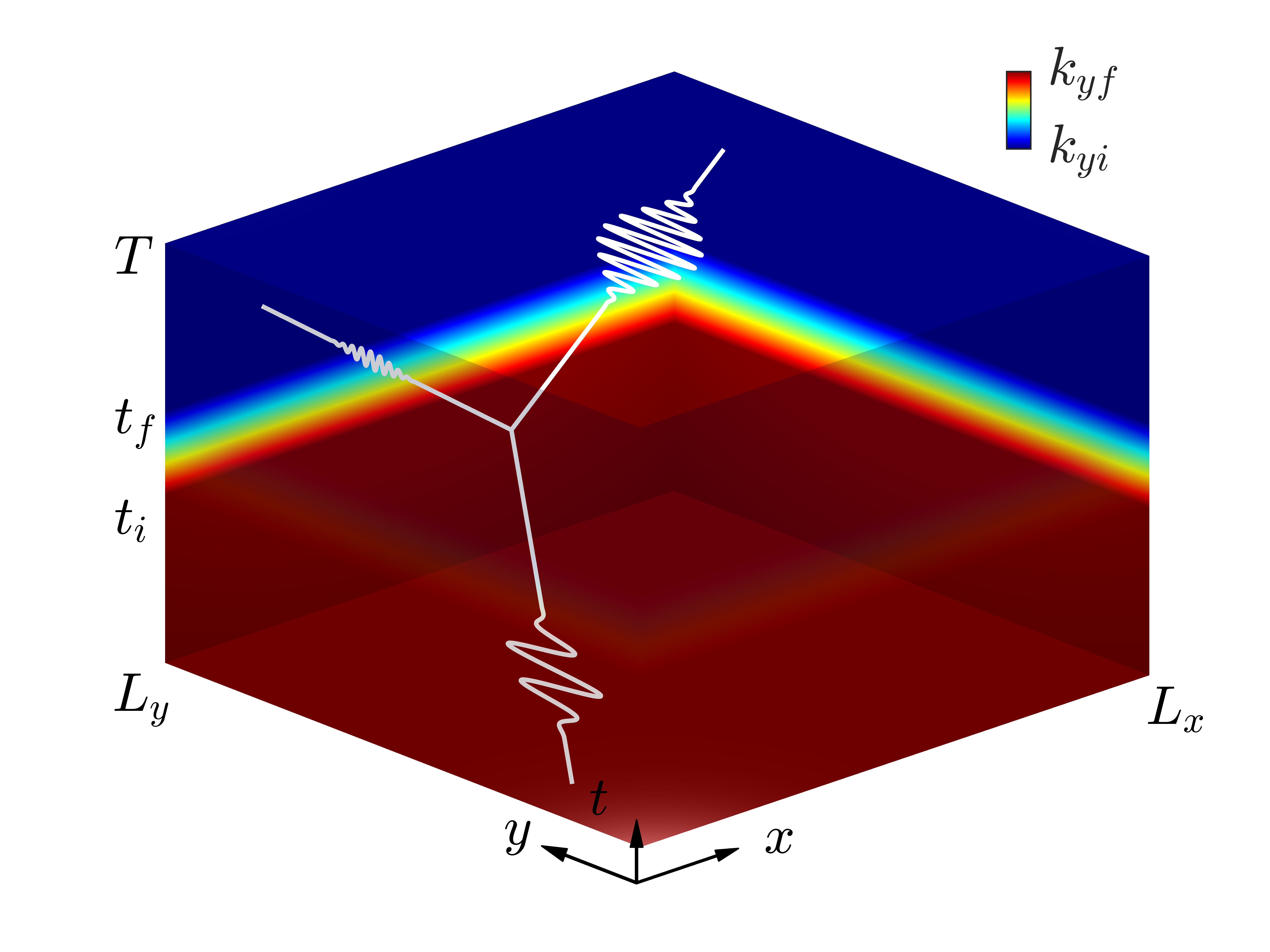}}
    \subfigure[]{\includegraphics[width=0.34\textwidth]{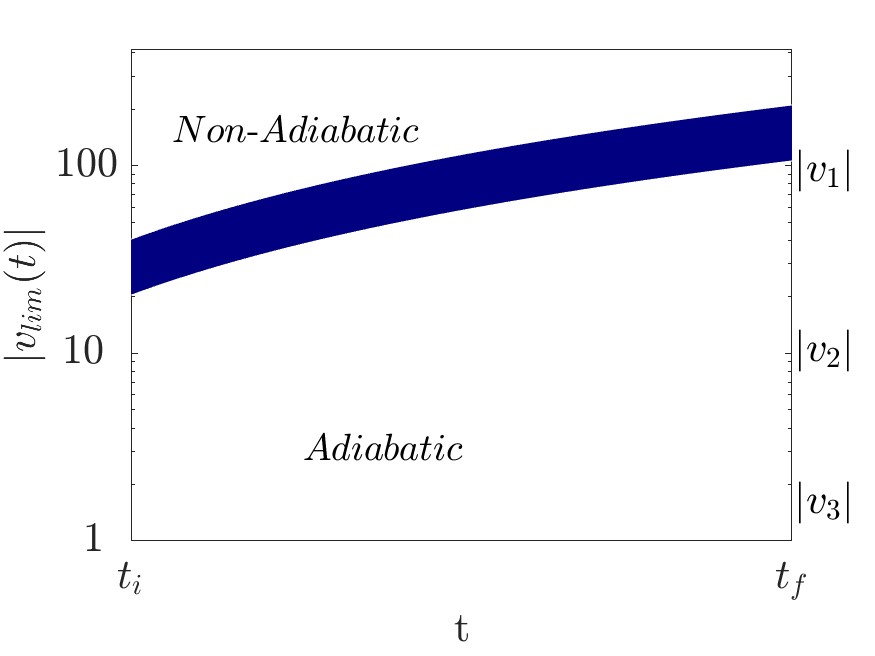}}
    \caption{(a) Schematic of the stiffness-modulated 2D spring-mass lattice, (b) along with the visual representation of the scattering phenomena taking place during time modulation. The colors represent time modulation, while the white curve illustrates an incident wavepacket which undergoes frequency modulation and steering. (c) Limiting condition for the modulation velocity, which delineates the transition between adiabatic and non-adiabatic transformations upon varying time. The three modulation velocities employed in the simulations are indicated alongside the diagram.}
    \label{fig:2D}
\end{figure}
The discussion is now focused on the elastic lattice displayed in Fig. \ref{fig:2D}(a), which is made of $60\times 60$ masses equally spaced by the lattice constant $d$, and connected to the nearest neighbors through linear springs $k_x$ and $k_y$. For simplicity and without any loss of generality, $k_y\left(t\right)$ accommodates the temporal degree of freedom. By following the same line of work, we consider a linear time varying function for $k_y\left(t\right)=k_{yi}+v\left(t-t_i\right)$, capable of driving the stiffness value from $k_{yi}$ to $k_{yf}$ in a time interval $\Delta t = t_f - t_i$ with modulation velocity $v$. The concept is elucidated in Fig. \ref{fig:2D}(b): an impinging wave that propagates across a temporal discontinuity experiences a frequency transformation and, provided that the modulation is adiabatic, the process occurs with negligible energy leak toward the back-propagating state. 
We hereafter demonstrate that such a frequency transformation is accompanied by a time-dependent curvature of wave motion, which serves as a mechanism to functionally control and guide waves.

We start the analysis from the time-dependent elastodynamic equation, which describes out-of-plane motion of the mass element sitting in position $\bm{r}=\left(n,m\right)$:
\begin{equation}
    m\ddot{u}_{n,m} + 2\left(k_x + k_y\right)u_{n,m} - k_x\left(u_{n+1,m}+u_{n-1,m}\right) - k_y\left(u_{n,m+1}+u_{n,m-1}\right) = 0
    \label{eq:dyn3}
\end{equation}
where to ease notation, the time dependence of $k_y\left(t\right)$ is implicitly assumed.
Now, wave propagation in such a system is investigated via a dispersion analysis of the unit cell with implied Floquet-Bloch boundary conditions, whereby the displacement $u_{n,m}\left(t\right)$ is described via complex exponential functions $u_{n,m} = \hat{u}e^{j\bm{\mu}\cdot\bf{r}}$, where $\bm{\mu}=\left(\mu_x,\mu_y\right)$ is the dimensionless wavevector. After a few mathematical steps we get to the following form:
\begin{equation}
    |\hat{\bm{\Psi}}\rangle,t=H\left(\mu,t\right)|\hat{\bm{\Psi}}\rangle\hspace{1cm} |\hat{\bm{\Psi}}\rangle=\begin{pmatrix}
        \hat{p}\\
        \hat{u}
        \end{pmatrix}\hspace{1cm}
    H\left(\bm{\mu},t\right)=
    \begin{bmatrix}
        0&-2\left[k_x(1-\cos\mu_x)+k_y(1-\cos\mu_y)\right]\\[5pt]
        \displaystyle\frac{1}{m}&0
    \end{bmatrix}
    \label{eq:sys2}
\end{equation}
which provide suitable description of the quasi-static evolution of two propagating eigenstates relative to an impinging wave $|\hat{\bm{\Psi}}\rangle=|\bm{\psi}^R\rangle{\rm e}^{j\omega t}$ with wavevector $\bm{\mu}$. 
The eigenstates, solution of the quasi-static eigenvalue problem $H\left(\bm{\mu},t\right)|\bm{\psi}^R\rangle=j\omega|\bm{\psi}^R\rangle$ are:
\begin{equation}
\begin{aligned}
        &\hspace{3cm}\omega_{1,2} = \pm\sqrt{\frac{2}{m}\left(k_x\left(1-\cos{\mu_x}\right)+k_{y}\left(1-\cos{\mu_y}\right)\right)}\\[10pt]
        &|\bm{\psi_1}^R\rangle = 
        \begin{pmatrix}
            j\sqrt{m\left(k_x\left(\cos{\mu_x}-1\right)+k_y\left(\cos{\mu_y}-1\right)\right)}\\[6pt]
            \displaystyle\frac{\sqrt{2}}{2}
        \end{pmatrix}\hspace{0.5cm}
        \langle\bm{\psi_1}^L| = \begin{pmatrix}
            \displaystyle\frac{1}{j\sqrt{4m\left(k_x\left(\cos{\mu_x}-1\right) + k_y\left(\cos{\mu_y}-1\right)\right)}}&
            \displaystyle\frac{\sqrt{2}}{2}
        \end{pmatrix}
        \\[10pt]
        &|\bm{\psi_2}^R\rangle = \begin{pmatrix}
            -j\sqrt{m\left(k_x\left(\cos{\mu_x}-1\right)+k_y\left(\cos{\mu_y}-1\right)\right)}\\[6pt]
            \displaystyle\frac{\sqrt{2}}{2}
        \end{pmatrix}\hspace{0.5cm}
        \langle\bm{\psi_2}^L| = \begin{pmatrix}
            -\displaystyle\frac{1}{j\sqrt{4m\left(k_x\left(\cos{\mu_x}-1\right) + k_y\left(\cos{\mu_y}-1\right)\right)}}&
            \displaystyle\frac{\sqrt{2}}{2}
        \end{pmatrix}
        \end{aligned}
        \label{eq:eigVec2}
\end{equation}
with suitable normalization $\langle\bm{\psi}_{1,2}^L|\bm{\psi}_{1,2}^R\rangle=\delta_{1,2}$. Finally, analytical forms of the group velocity $\bm{c}_g=\nabla\omega\left(\bm{\mu}\right)$ for the above eigenstates can be easily deduced:
\begin{figure}
    \subfigure[\label{profiles}]{\includegraphics[width=0.39\textwidth]{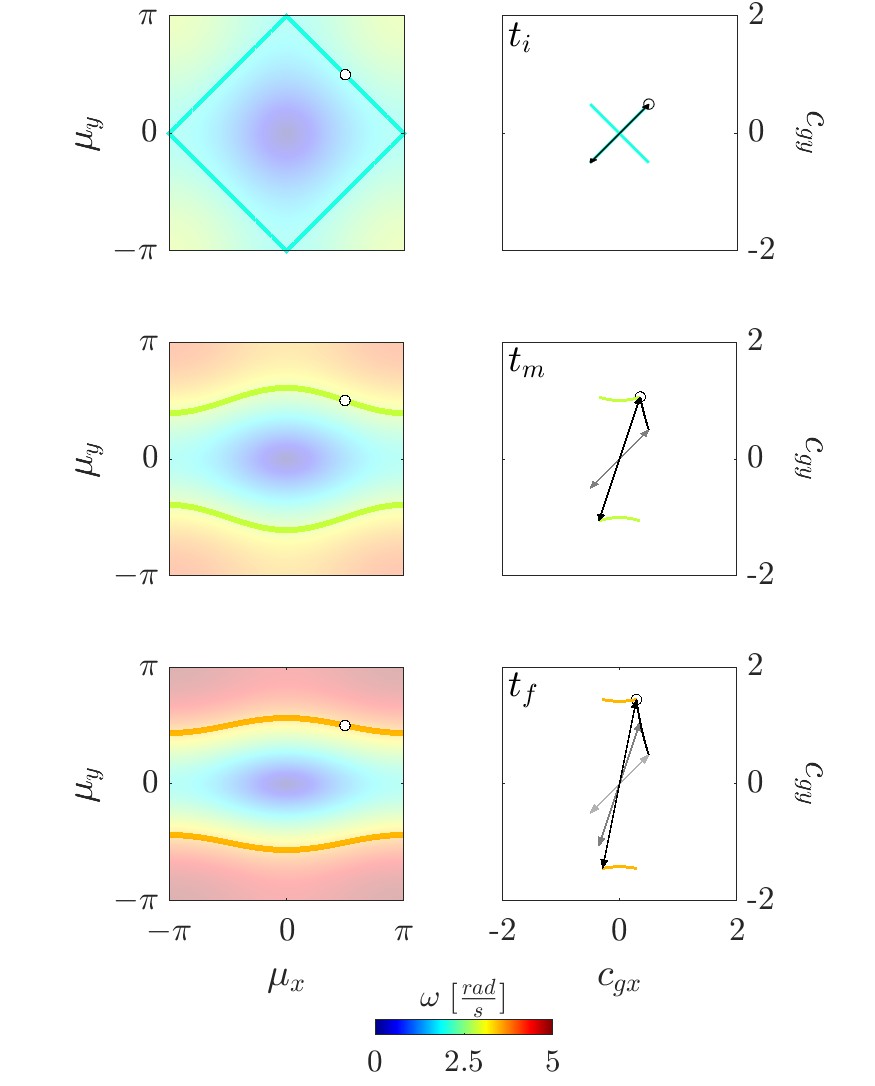}}\hspace{-0.1cm}
    \subfigure[\label{time}]{\includegraphics[width=0.60\textwidth]{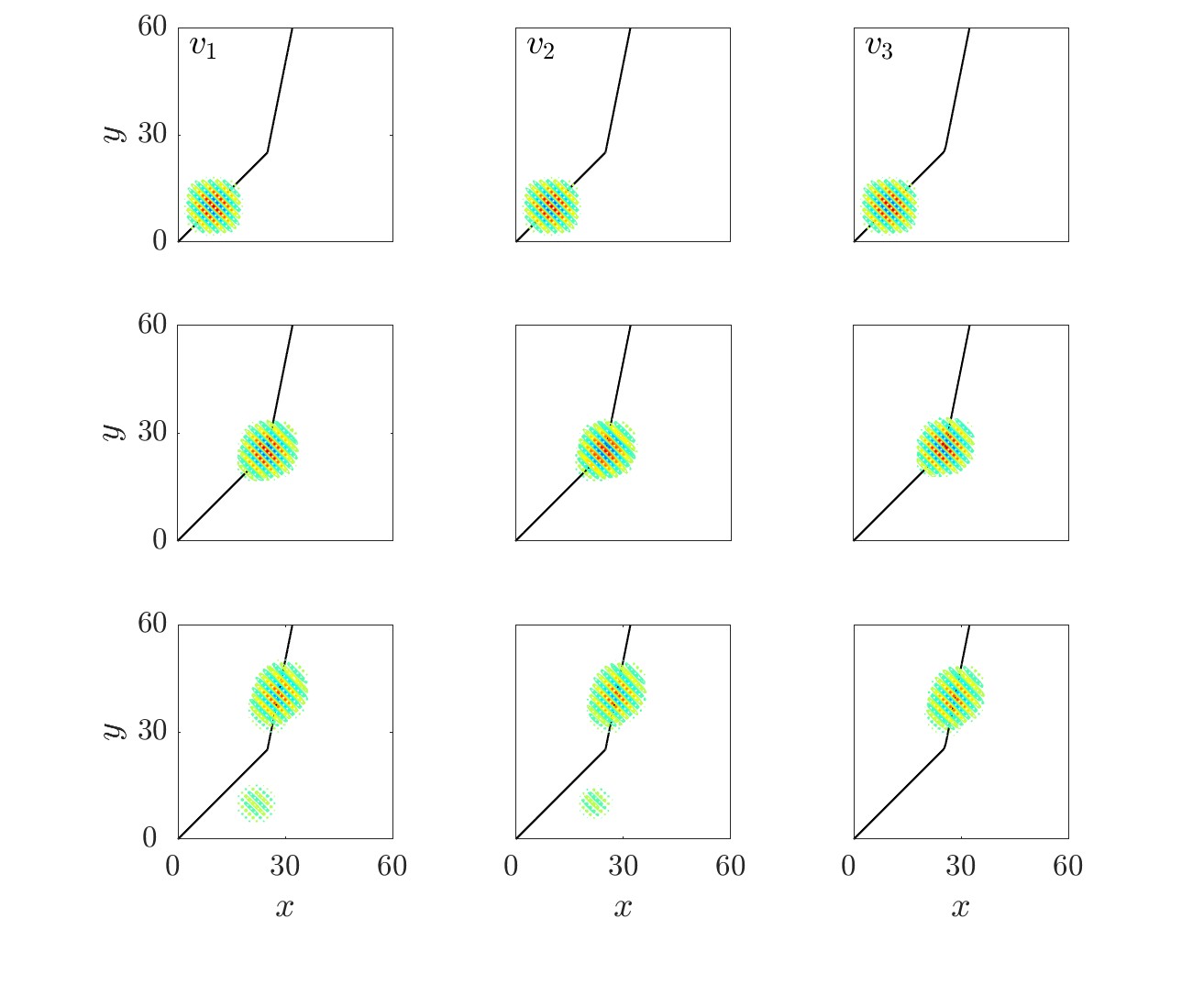}}
    \subfigure[\label{2DfastSpectr}]{\includegraphics[width=0.28\textwidth]{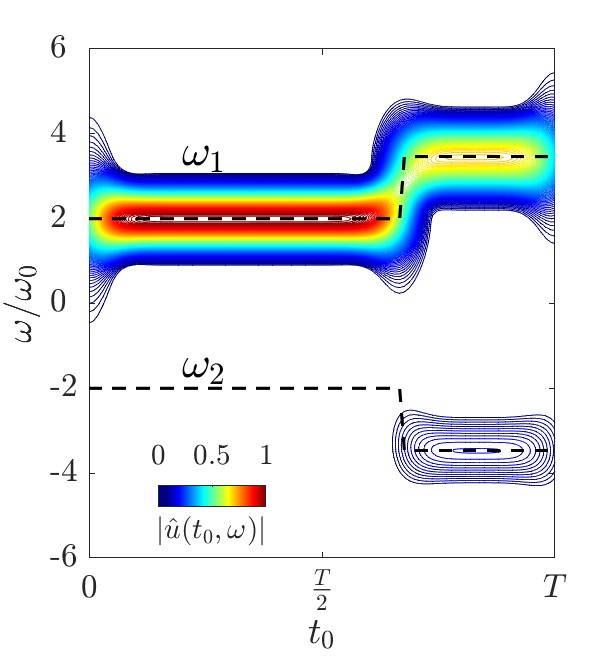}}\hspace{0.5cm}
    \subfigure[\label{2DmediumSpectr}]{\includegraphics[width=0.28\textwidth]{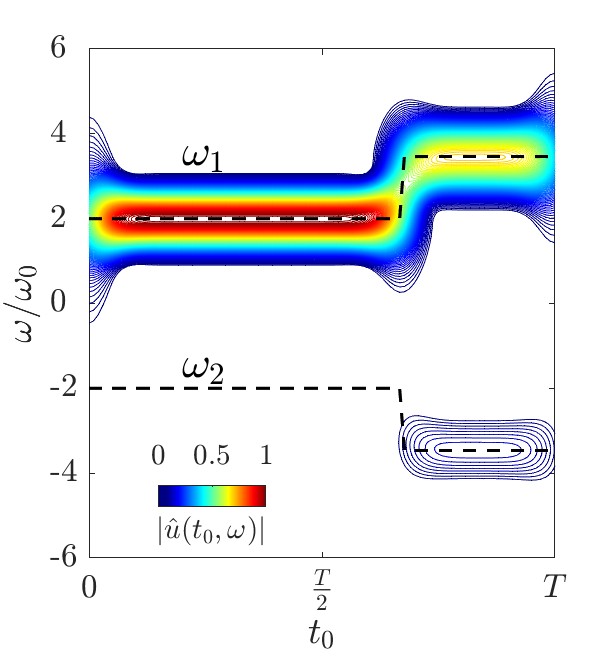}}\hspace{0.5cm}
    \subfigure[\label{2DslowSpectr}]{\includegraphics[width=0.28\textwidth]{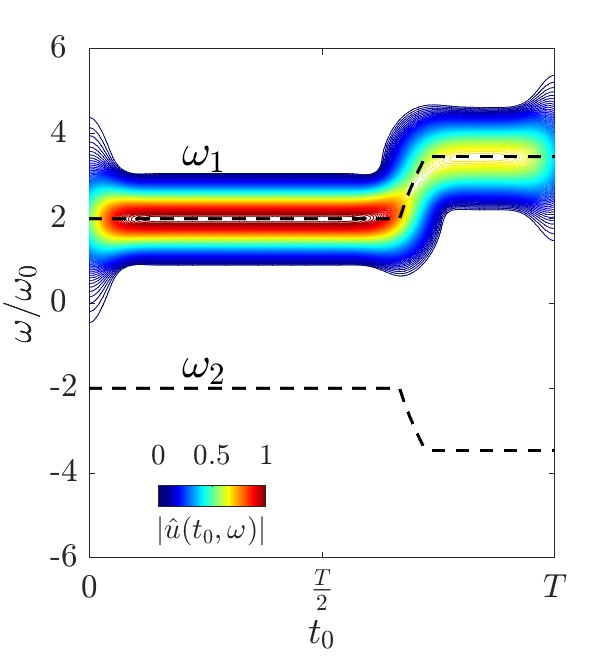}}
    \caption{(a) Dispersion relation (left) and corresponding group velocity profiles (right) for three different  time instants: before the modulation $t = t_i$, during time modulation $t_m = (t_i+t_f)/2$ and after time modulation $t = t_f$. The white dot represents the central wavevector content of the incident wavepacket employed for numerical simulations. (b) Time histories relative to the three modulation velocities $v_1 = 93$, $v_2 = 10$, and $v_3 = 1.6$. The black curve represents the desired trajectory predicted through the group velocity profile. (c-e) Spectrograms to elucidate how the frequency content within the lattice varies in time for modulation velocities (c) $v_1$, (d) $v_2$, and (e) $v_3$. The colored contours are relative to the Fourier transformed version of the displacement field, while the expected evolution of the states is represented with black dashed lines.}
    \label{fig:Profiles}
\end{figure}
\begin{equation}
        \bm{c}_{g} = \left(\frac{k_x d\sin{\mu_x}}{\sqrt{2m\left(k_x\left(1-\cos{\mu_x}\right)+k_y\left(1-\cos{\mu_y}\right)\right)}}\hspace{0.25cm}\frac{k_y d\sin{\mu_y}}{\sqrt{2m\left(k_x\left(1-\cos{\mu_x}\right)+k_y\left(1-\cos{\mu_y}\right)\right)}}\right)
    \label{eq:GroupSpeed}
\end{equation}
where counter propagating states have opposite velocity sign $\bm{c}_{g\;1,2}=\pm\bm{c}_{g}$. The direction of propagation $\phi$ follows:
\begin{equation}
    \tan{\phi(t)} = \frac{c_{gy}}{c_{gx}} = \frac{k_y}{k_x}\frac{\sin{\mu_y}}{\sin{\mu_x}}
    \label{eq:phi}
\end{equation}
Interestingly, an incident state with arbitrary wavevector $\bm{\mu}$ propagates along the direction $\phi$ defined by the group velocity $\bm{c}_g\left(\bm{\mu}\right)$, which is tailorable with $k_y\left(t\right)$. In other words, the analytical forms for $\omega_{1,2}$ and $\phi\left(\bm{\mu}\right)$ delineate both frequency conversion and curvature of wave motion induced by the stiffness modulation which, in turn, can be tailored to functionally steer waves via adiabatic and nonadiabatic transformations. 

We now discuss on the dynamics of smooth temporal modulations applied to the 2D lattice, and the relative transformation that takes place when an incident wave packet propagates in such a system. First and foremost, Eq. \ref{eq:sys2} have the same form of Eq. \ref{eq:sys} and,
as such, for any impinging wavevector $\bm{\mu}$, there are two counter propagating states that populate the lattice. It is straightforward to conclude that Eq. \ref{eq:limit} can be employed to assess adiabaticity of the modulation, and gives the following condition for the velocity in order for a propagating state $|\bm{\psi_1}^R\rangle{\rm e}^{j\omega_1t}$ not to leak energy toward $|\bm{\psi_2}^R\rangle{\rm e}^{j\omega_2t}$: 
\begin{equation}
    |v_{lim}| <<
    16\displaystyle\frac{\sqrt{\left|k_x\sin^2{\displaystyle\frac{\mu_x}{2}}-k_y\sin^2{\displaystyle\frac{\mu_y}{2}}\right|}}{\sqrt{m}\sin^2{\displaystyle\frac{\mu_y}{2}}}
    \label{eq:limSpeed2D}
\end{equation}
In analogy to the procedure employed in the previous section, Fig. \ref{fig:2D}(c) illustrates the limiting condition for the speed of modulation to be considered adiabatic.
This condition is valid for the entire wavenumber content of the incident wave packet, which consists of a prescribed displacement with central wavnumber $\bm{\mu}= [0.5\pi, 0.5\pi]$ and number of periods $n_x=n_y=5$.
To get to Fig. \ref{fig:2D}(c), the wavenumber dimension is eliminated by taking the minimum $\displaystyle\min_{D}\{ v_{lim}\left(\bm{\mu},t\right)\}$ and maximum $\displaystyle\max_{D}\{ v_{lim}\left(\bm{\mu},t\right)\}$ values within the domain $D=\mu_x\times\mu_y=\left[0.3\pi,0.7\pi\right]\times\left[0.3\pi,0.7\pi\right]$, which allowed us to include in the analysis all the relevant spectral content of the wavepacket.

Three distinct modulation velocities $v_1 = 93$, $v_2 = 10$, and $v_3 = 1.6$ are probed to qualify consistency of Eq. \ref{eq:limSpeed2D} and Fig. \ref{fig:2D}(c) via numerical simulation: it is expected that these values, defined across the limiting condition, allow us to shed light on the energy leaks which manifests in time-modulated lattices with concurrent frequency conversion and wave steering. We firstly show dispersion relations and group velocity profiles in Fig. \ref{fig:Profiles}(a) for consecutive time instants $t_i$, $t_m=\left(t_i+t_f\right)/2$ and $t_f$. We focus our attention on the central wavevector content $\bm{\mu}=\left(0.5\pi,0.5\pi\right)$ of the incident wave packet (highlighted with a white dot in the dispersion diagram). For such a point in reciprocal space, the corresponding group velocities are reported alongside the dispersion, with emphasis on how the directionality evolves over time. It is expected that the temporal evolution of a narrowband wavepacket does follow the directionality dictated by \ref{fig:Profiles}(a). Results from time simulations follow. 

\begin{figure}
    \centering
    \subfigure[\label{time}]{\includegraphics[width=.83\textwidth]{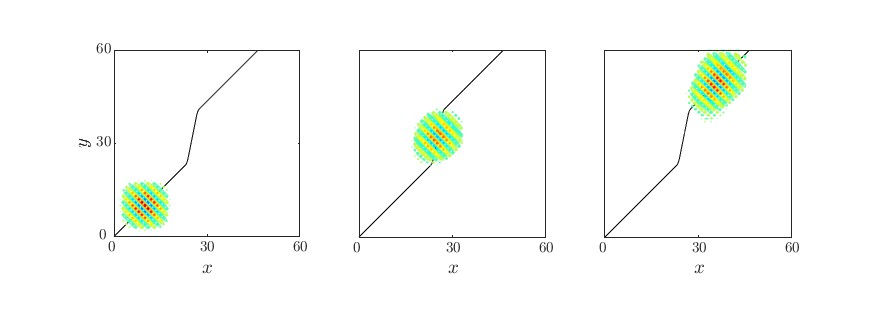}}\\
    \hspace{-0.5cm}
    \subfigure[\label{time}]{\includegraphics[width=.4\textwidth]{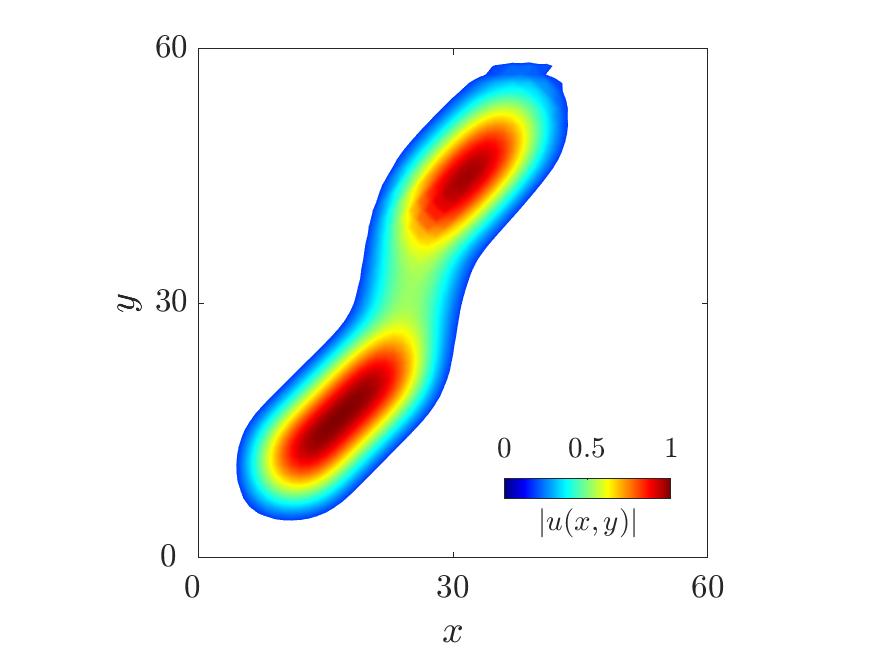}}\hspace{-1cm}
    \subfigure[\label{time}]{\includegraphics[width=.4\textwidth]{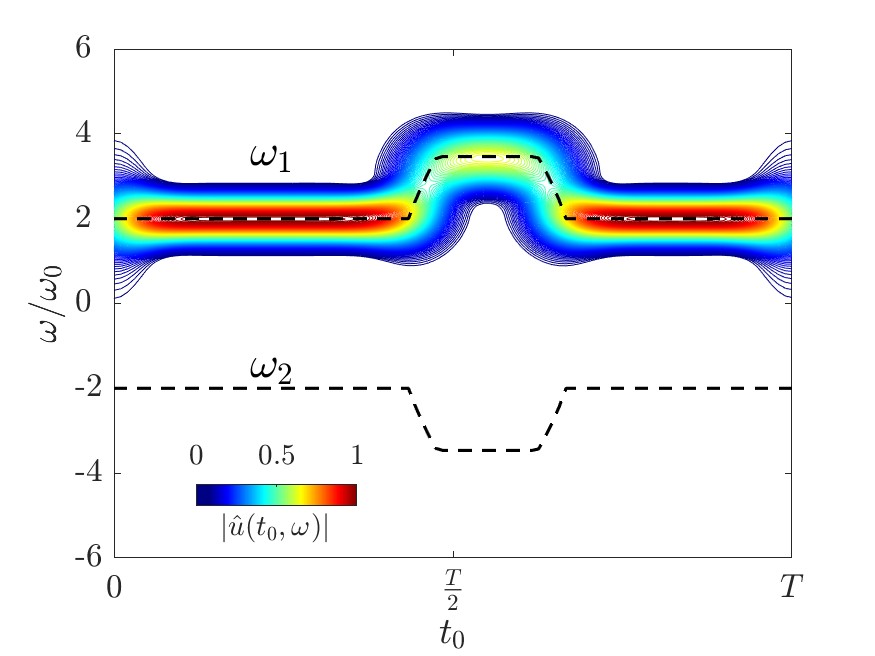}}
    \caption{(a) Snapshots of the displacement field for a 2D lattice modulated in time via three consecutive modulation levels (colored contours). The displacement field is superimposed to the desired trajectory (black line). (b) RMS of the displacement field in time. (c) Time spectrogram for the time evolution of the frequency content in time (colored contours) and expected evolution of the states (black dashed lines).}
    \label{fig:ComplexDyn}
\end{figure}

Three relevant snapshots prior, during, and after the modulation are illustrated in Fig. \ref{fig:Profiles}(b) for the three aforementioned modulation velocities. For comparison, the expected trajectory is predicted from integration of the group velocities in time and displayed in black underneath the displacement fields. We observe that all configurations are capable of curving the wave packet by the desired angle $\Delta\phi$ and, hence, are consistent with the group velocities illustrated in Fig \ref{fig:Profiles}(a). In case of fast modulation $v_1$, there is a relevant wave packet that separates from the incident state $\omega_1$, which is blueprint of energy leak across the temporal discontinuity. The reflected wave packet effectively embodies the energy transferred to state $\omega_2$ via non-adiabatic passage, and propagates with opposite direction as compared to $\omega_1$. To decrease the energy leak, a sufficiently slow velocity $v$ is to be employed. Configurations with $v=v_2$ and $v=v_3$ exhibit greater degree of adiabaticity and, hence, the reflection is minimal or absent. Animations are reported in the supplementary material.

To further verify the above considerations, the frequency spectrograms relative to the time histories are displayed in Fig. \ref{fig:Profiles} (c-e). In analogy to 1D lattices, the wavenumber domain is eliminated by taking the $L^2$ norm limited to $\mu_x>0$ and $\mu_y>0$. The results are consistent to the theory: fast and intermediate modulations (see Fig. \ref{fig:Profiles}(c-d)) are not adiabatic and, as such, there is energy transfer from the positive toward the negative frequencies. In contrast, sufficiently slow modulations (Fig. \ref{fig:Profiles}(e)) induce adiabatic transformations whereby the energy content represented in reciprocal space does not leave the incident state. Hence, in physical space, the trajectory of an incident wave-packet can be curved without undesired reflections. 

Finally, we design a more complex adiabatic law for $k_y$, which consists of three consecutive stiffness levels $k_y=1$, $k_y=10$, and back to $k_y=1$. Fig. \ref{fig:ComplexDyn}(a) illustrates temporal snapshots of the wave packet that, due to stiffness modulation, is capable of steering at will (animations are reported in the supplementary material). The curvature of wave motion is further highlighted by taking the $RMS$ along the temporal dimension (Fig. \ref{fig:ComplexDyn}(b)), whereby the lower amplitude in the central part is attributed both to a lower amplitude of wave motion and greater velocity of propagation.  To conclude the paper, we demonstrate adiabaticity of this transformation which, according to the spectrogram in Fig. \ref{fig:ComplexDyn}(c) occurs without energy scattering toward counter propagating states.

\section{conclusions}
In this paper, we explored the dynamics of slow temporal modulations in the context of 1D and 2D lattices. The behavior of such systems is driven by the time-evolution of the underlying dispersion properties which, under quasi-static conditions, dictate both frequency conversion and directionality of wave motion. 
By way of the adiabatic theorem, we define a condition in order for the transformation to occur without energy leakage, which is responsible for undesired reflections. The concept can be easily extended to physical systems with tunable dispersion relations, such as beams with piezoelectric materials in mechanics \cite{marconi2020experimental,xia2021experimental}, and materials with tunable permittivity \cite{pacheco2020temporal}. Although based on a very different phenomena, we herein propose a waveguiding mechanism that allows an elastic signal to be sent from an emitter to a receiver, similarly (but conceptually very different) to what has been observed in many studies concerning quantum valley Hall (QVH) and quantum spin Hall (QSH)-based waveguides.

\bibliography{References}

\end{document}